\documentclass[manuscript]{emulateapj}

\slugcomment{}

\shorttitle{}
\shortauthors{Cappelluti et al.}

\begin{document}


\title{PROBING LARGE SCALE COHERENCE BETWEEN {\em SPITZER} IR AND {\em CHANDRA} X-RAY SOURCE-SUBTRACTED COSMIC BACKGROUNDS}


\author{N. Cappelluti\altaffilmark{1,2}, R. Arendt\altaffilmark{3,4}, A. Kashlinsky\altaffilmark{4,5}, Y. Li\altaffilmark{6},
G. Hasinger\altaffilmark{6}, K. Helgason\altaffilmark{7} M. Urry\altaffilmark{1,2}, P. Natarajan\altaffilmark{8},A. Finoguenov\altaffilmark{9,10}}


\altaffiltext{1}{Yale Center for Astronomy and Astrophysics, P.O. Box 208120, New Haven, CT 06520.}
\altaffiltext{2}{Department of Physics, Yale University, P.O. Box 208120, New Haven, CT 06520.}
\altaffiltext{3}{University of Maryland, Baltimore County, 1000 Hilltop Circle, Baltimore, MD 21250, USA}
\altaffiltext{4}{Observational Cosmology Laboratory, Code 665, Goddard Space Flight Center, Greenbelt MD 20771, USA}
\altaffiltext{5}{SSAI, 10210 Greenbelt Road, Suite 600, Lanham, MD 20706, USA}
\altaffiltext{6}{Institute for Astronomy, 2680 Woodlawn Drive, University of Hawaii, Honolulu, HI 96822, USA}
\altaffiltext{7}{Department of Astronomy, Yale University, PO Box 208101, New Haven, CT 06520.}
\altaffiltext{8}{Max Planck Institute for Astrophysics, Karl-Schwarzschild-Str. 1, D-85748 Garching, Germany}
\altaffiltext{9}{Max-Planck-Institut f\"ur extraterrestrische Physik, Postfach 1312, 85741, Garching bei M�nchen, Germany}
\altaffiltext{10}{Department of Physics, University of Helsinki, Gustaf H\"allstr\"omin katu 2a, FI-00014 Helsinki, Finland}


\begin{abstract}
We present new measurements of the large scale clustering component of the cross-power spectra of the source-subtracted {\em Spitzer}-IRAC Cosmic Infrared Background (CIB) and {\em Chandra}-ACIS Cosmic X-ray Background (CXB) surface brightness fluctuations.
 Our investigation uses data from the {\em Chandra} Deep Field South (CDFS), {\em Hubble} Deep Field North (HDFN), EGS/AEGIS field and UDS/SXDF surveys, 
comprising 1160 Spitzer hours and $\sim$ 12 Ms of {\em Chandra} data collected over a total area of 0.3 deg$^2$. We report the first ($>$5$\sigma$) detection of a cross-power signal on large angular scales 
$>\,$20$\arcsec$ between [0.5-2]~keV and the 3.6 and 4.5$\mu$m bands, at $\sim$5$\sigma$ and 6.3$\sigma$ significance, respectively. 
The correlation with harder X-ray bands is marginally significant. 
Comparing the new observations with existing models for the contribution of the known unmasked 
source population at $z<$7, we find an excess of about an order of magnitude at 5$\sigma$ confidence.
We discuss possible interpretations for the origin of this excess in terms of the contribution from accreting early black holes, including both direct collapse 
black holes and primordial black holes, as well as from scattering in the interstellar medium and intra-halo light. 

\end{abstract}





\section{Introduction}

The Cosmic Infrared Background (CIB) is produced from the integrated radiation resulting from stars, accretion and 
dust reprocessing, from the epoch of the last scattering to the present. Although most of the CIB flux has been resolved into discrete sources,
a sizable fraction of them are inaccessible for telescopic follow-up studies either because they are intrinsically faint or very distant. Of particular interest 
is the study of the contribution to the CIB from sources at the epoch of the first stars and Black Holes (BHs). Current understanding of structure formation
suggests that these first UV-bright objects form between $z = 15 - 25$, and their radiation would be redshifted to the infrared today. Therefore, the properties
of the CIB---in particular, the surface brightness fluctuations---offer a new window to access these high-redshift sources by studying the residuals after the removal of known sources. 

An excess,  of  about a factor $>20$, on scales larger 
than $\sim$30$\arcsec$ with respect to known, z$<$6, populations was detected by \citet{kamm1,kamm3,k12}, and confirmed by \citet{c12b}, with {\em Spitzer}  after removing sources to m$_{AB}$=24-25 at 3.6$\mu$m and 4.5$\mu$m, and  \citet{m11,seo} with {\em AKARI} below  m$_{AB}$=23-24 at 2.4$\mu$m, 3.2 $\mu$m and 4.1$\mu$m. Its origin is debated: it can be attributed entirely to high-redshift sources \citep{kamm3,k12,yue13} or to diffuse intra-halo light 
around galaxies at $z=1-5$ \citep[][]{c12,zemcov} and \citep[for a review]{k17}. 
 At shorter wavelengths, not directly relevant to this study, the situation is less clear with conflicting measurements from 2MASS \citep{k02,o3}, NICMOS \citep{t07,t072} and CIBER \citep{zemcov} as discussed in detail in Sec. 2.1.2 of \citet{k15}.

\citet{cap2013} measured a statistically significant cross-power spectrum between the source-subtracted {\em Spitzer} CIB and {\em Chandra} CXB [0.5-2]~keV 
fluctuations in the Extended Groth Strip (EGS), suggesting that sources responsible for the CIB excess share the same environment with, or are, accreting BHs. 
This result was confirmed by \citet{mw16}. 
Neither study was able to probe with high significance the cross-power at the largest scales (i.e., $>$20-30$\arcsec$) arising from clustering. \citet{h14} showed that known source populations alone (X-ray binaries, AGN and hot gas) are not sufficient to account for the tentative large scale component seen in the cross-power. 
\citet{yue13} interpreted this 
excess CIB power and the CIB-CXB coherence as arising from a population of Direct Collapse Black Holes 
\citep[DCBH, see e.g.][and refs. therein]{ln2, Lodato_Natarajan_2006} at z$>$12. 
Alternatively, \citet{k16} suggested that the measured CIB fluctuations could be explained naturally if LIGO events arise from primordial BHs 
(of $\sim$20-40 M$_\odot$) making up the dark matter \citep{bird},

Here, we present the first measurement of 
the clustering component in the source-subtracted CIB versus CXB fluctuations cross-power spectra, between four photometric band pairs, by combining the deepest {\em Spitzer} and {\em Chandra} observations available to date. 

\section{ Datasets and map production} \label{sect1}
\subsection{Chandra X-ray data }

\begin{table*}[!t]
\begin{center}
\tiny
\caption{X-ray map properties\label{tbl-1}}

\begin{tabular}{|llllllllllr|}
\hline
 \multicolumn{1}{|c}{Field} &
 \multicolumn{1}{c}{FoV} &
  \multicolumn{1}{c}{Pixel scale} &
 \multicolumn{1}{c}{$\sum$Exp} &
    \multicolumn{1}{c}{$\langle$Exp$\rangle$} &

  \multicolumn{1}{c}{N$_{ph,tot 0.5-2}$} &
  \multicolumn{1}{c}{N$_{ph,Astr. (0.5-2)}$} &
  \multicolumn{1}{c}{N$_{ph,tot (2-7)}$} &
  \multicolumn{1}{c}{N$_{ph,Astr. (2-7)}$} &
  \multicolumn{1}{c}{\%$_{mask}$} &
  \multicolumn{1}{l|}{Catalog} \\
	& arcmin & arcsec  & Ms & Ms && & & & & \\
	\hline
  HDFN & 9.6$\arcmin\times$5.1$\arcmin$ & 0.6\arcsec & 1.79 &1.62&  91904 & 55956 & 187911 & 99693 &38\%&\citet{alex}\\
  CDFS & 12.6$\arcmin\times$9.5$\arcmin$ & 0.6\arcsec &6.67 &5.96& 760651 & 110698 & 1943340 & 130973 &40\%&\citet{luo}\\
  EGS & 45.2$\arcmin\times$9.5$\arcmin$ & 1.2\arcsec & 2.25& 0.75 & 89484 &  49832 & 232297 & 118227 &32\%&\citet{goulding}\\
  UDS & 20.4$\arcmin\times$20.4$\arcmin$ & 1.2\arcsec&1.19 & 0.43& 59995 & 28792 & 134819 & 66700 & 36\%&\citet{koc} \\
\hline
\end{tabular}

\end{center}
\end{table*}
\begin{table*}[!th]
\begin{center}
\caption{\label{tbl-2} 
Cross-power-spectrum amplitude on $>$20\arcsec in units of 10$^{-19}$ erg s$^{-1}$ cm$^{-2}$ nW m$^{-2}$ sr$^{-2}$ }

\begin{tabular}{|lrrrr|}
\hline
& & $2\pi/q >20\arcsec$  &  & \\
\hline
  \multicolumn{1}{|c}{Field} &
  \multicolumn{1}{c}{3.6 $\mu$m vs [0.5-2]} &
  \multicolumn{1}{c}{4.5 $\mu$m vs [0.5-2]} &
  \multicolumn{1}{c}{3.6 $\mu$m vs [2-7]} &
  \multicolumn{1}{c|}{4.5 $\mu$m vs [2-7]} \\
\hline
 HDFN & 2.45$\pm{2.16}$ & 0.05$\pm{1.54}$ & -18.62$\pm{13.40}$ & -6.64$\pm{10.3}$ \\
  CDFS & 3.41$\pm{1.00}$ & 2.13$\pm{0.85}$& 9.00$\pm{6.91}$ & -0.38$\pm{6.47}$ \\
  EGS & 1.53$\pm{0.46}$ & 1.98$\pm{0.39}$ & 2.56$\pm{3.79}$ & 4.77$\pm{3.21}$ \\
  UDS & 2.04$\pm{0.94}$ & 1.57$\pm{0.61}$ & -7.50$\pm{5.87}$ & 2.04$\pm{3.35}$ \\
   STACK & 1.87$\pm{0.37}$& 1.91$\pm{0.34}$& 2.73$\pm{3.22}$& 3.00$\pm{2.76}$ \\
\hline  	

\end{tabular}
\end{center}
\end{table*}

The X-ray data are from the deep Chandra ACIS-I AEGIS survey \citep[EGS,][]{goulding,nandra}, the UDS-SXDF field 
\citep{koc}, the Chandra Deep Field South \citep{luo} and the Hubble Deep Field North \citep{alex}. 
In total we used 243 Chandra pointings yielding a total of 11.9 Ms of flare-cleaned data, over an area of 0.3 deg$^2$ (see Table \ref{tbl-1}). 
The data analysis methods used have been described in detail in \citet{cap2013} and \citet{mw16} except that here 
we apply a stricter rejection of flares \citep[see][]{cap2017} and use both the FAINT and VFAINT telemetry mode data. 
 
We created count maps in the [0.5-2]~keV and [2-7]~keV energy bands
from even (A) and odd (B) events to evaluate the noise floor for the cross-power (see below). 
We produced X-ray masks, M$_X$, with the catalogs listed in Table \ref{tbl-1}, by cutting circular regions of 7$\arcsec$ radius around each point-source: 
this removes $>$90\% of known X-ray source flux over the field. 

 We masked all the extended emission identified with groups and clusters of galaxies. 
Sensitivities and redshift ranges vary slightly from field to field and are discussed in \citet{fin15,erf13,fin10,erf14} for
CDFS, EGS, UDS and HDFN, respectively (see Fig. 3).

 Multiplying M$_X$ by the corresponding IR masks, M$_{IR}$, we obtain M$_{IR,X}$ \citep[and see below]{cap2013,kamm3,k12}.
In Table \ref{tbl-1} we summarize the number of X-ray counts in the maps after masking. The background was modeled with two components, 
one from particles/instrument and one from astrophysical sources. The first component, X$_p$, has been estimated with the ACIS-$stowed$ event 
files \citep[e.g.,][]{hm06,cap2013,cap2017} and the second, X$_{Cosm}$, by distributing remaining counts in the field according to an exposure map, E
\citep[for an extensive discussion see][]{cap2013,cap2017}; the mean X-ray map is $\langle X \rangle$=(X$_p$+X$_{Cosm}$)*M$_{IR,X}$.
The fluctuation map is then given as: $\delta F^i_{X_{raw}}=X^i/E^i-\langle X^i \rangle/E^i$ where X$^i$ are the X-ray counts at the position $i$. We weighted our maps to take vignetting into account and the resulting fluctuation map is $\delta F^i_{X}=\delta F^i_{X_{raw}}*E^i/\langle E \rangle$.
 
\subsection{Spitzer IR data}

The {\it Spitzer}/IRAC self-calibrated mosaics are the same data (program ID = 169, 194, 61041, 61042) that were analyzed in \citet{k12,kamm2} and described in detail in \citet{a10}.  We have reprocessed the observations with a new version of the self-calibration. Both epochs of the observations for each field were self-calibrated simultaneously using a new data 
model. The single field-of-view ``ultra-deep'' portion of the HDFN was still omitted, as well as the southern part of the HDFN,
 which was affected by an artifact.
 The new data model can be written as: $D^i = S^{\alpha} + F^{p,r} + F^{q},$ where $D^i$ are the data 
 in the $i$th pixel, 
 $S^{\alpha}$ is the sky
 intensity at position $\alpha$, $F^{p,r}$ is the ``fixed'' detector offset for each pixel $p$ for each group of frames (AOR) $r$, and $F^q$ is the ``variable'' detector offset as a 
 function of frame and output $q$ \citep[cf.][]{a10}. 

The new feature included in our analysis is the addition of the $r$ index which allows the fixed detector offset to vary in time. Previously, the data from each AOR were 
self-calibrated separately to derive the $F^p$ terms individually, and then the results were merged and remapped. Because of the changing zodiacal light between epochs, we had not 
been able to combine the observations. With the new self-calibration model, consistency of the derived sky and the $F^{p,r}$ and $F^q$ offsets is now built into the procedure. The resulting 
sky maps do not exhibit large scale variations caused by combining epochs. In addition, our enhanced procedure also slightly reduces small scale variations (noise) because the offsets are 
being determined relative to a sky that is the average of all the available data, not just the data from 1 AOR.
All maps were then clipped to the same shot noise level 
P$_{SN}$=50 and P$_{SN}$=30 nJy~nW/m$^2$/sr in 3.6$\mu$m and 4.5$\mu$m, respectively, or down to m$_{AB}\sim$24.8 mag.
X-ray maps astrometry has been matched to that of the IR maps with pixel scales of 0.6\arcsec~ for HDFN and CDFS and 1.2\arcsec~ for the  UDS and EGS.
\section{Fluctuation analysis}\label{sect2}

\begin{figure*}[!t]
\center
\includegraphics[width=\textwidth]{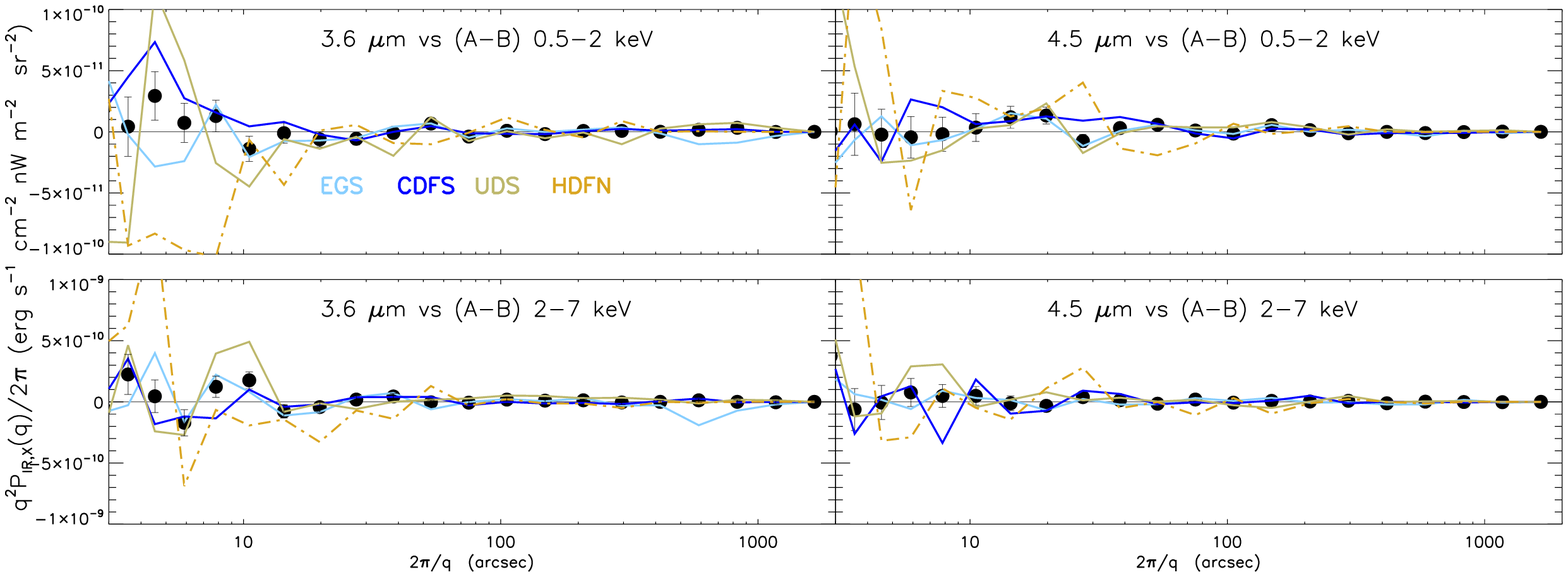}
\includegraphics[width=\textwidth]{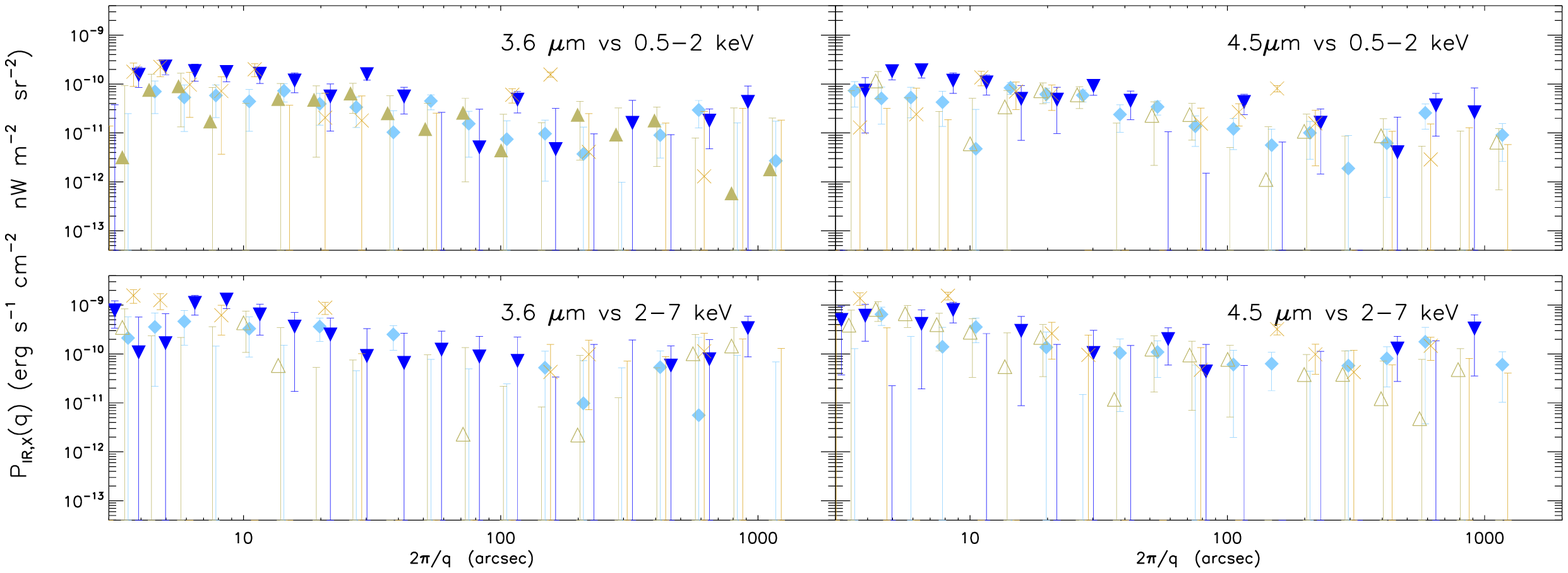}
\caption{\label{fig1}$Top~panels:$ The mean square fluctuation cross-power spectra between CIB fluctuations and CXB (A-B) maps as a function of the angular scale;
these indicate the level of systematic error in our analysis.
From $top~left$ to $bottom~right:$ 3.6~$\mu$m vs. 0.5-2~keV, 4.5~$\mu$m vs. 0.5-2~keV, 3.6~$\mu$m vs. 2-7~keV and 4.5$\mu$m vs. 2-7~keV, respectively. 
The noise is roughly 10 times lower for the soft X-ray band ($top panels$) compared to the hard X-ray band ($next two panels$). 
Colors refer to individual survey fields: 
$Cyan$ for EGS, $blue$ for CDFS, $light~brown$ for UDS and $light~orange$ for HDFN. The $black~filled~circles$ show the combined cross-power for all four fields.
$Bottom~panels:$ The same but between CIB  and CXB fluctuations maps in individual fields. Order is the same as in top panels. 
$Cyan~diamonds$ are EGS, $blue~down~facing~triangles$ are CDFS, $light~brown~triangles$ are UDS and $light~orange~crosses$ are HDFN.}
\end{figure*}

\begin{figure*}
\center
\includegraphics[width=\textwidth]{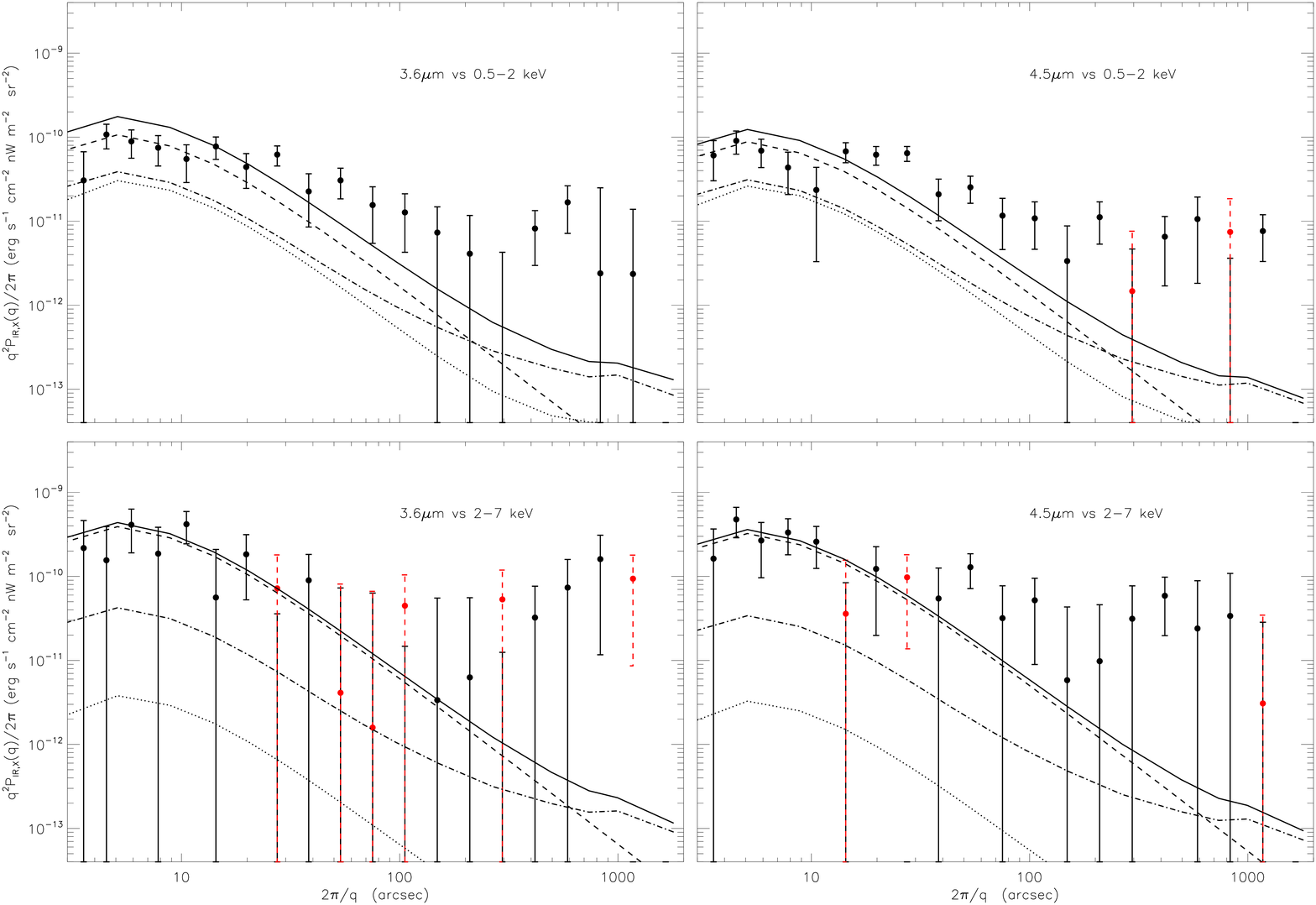}
\includegraphics[width=\textwidth]{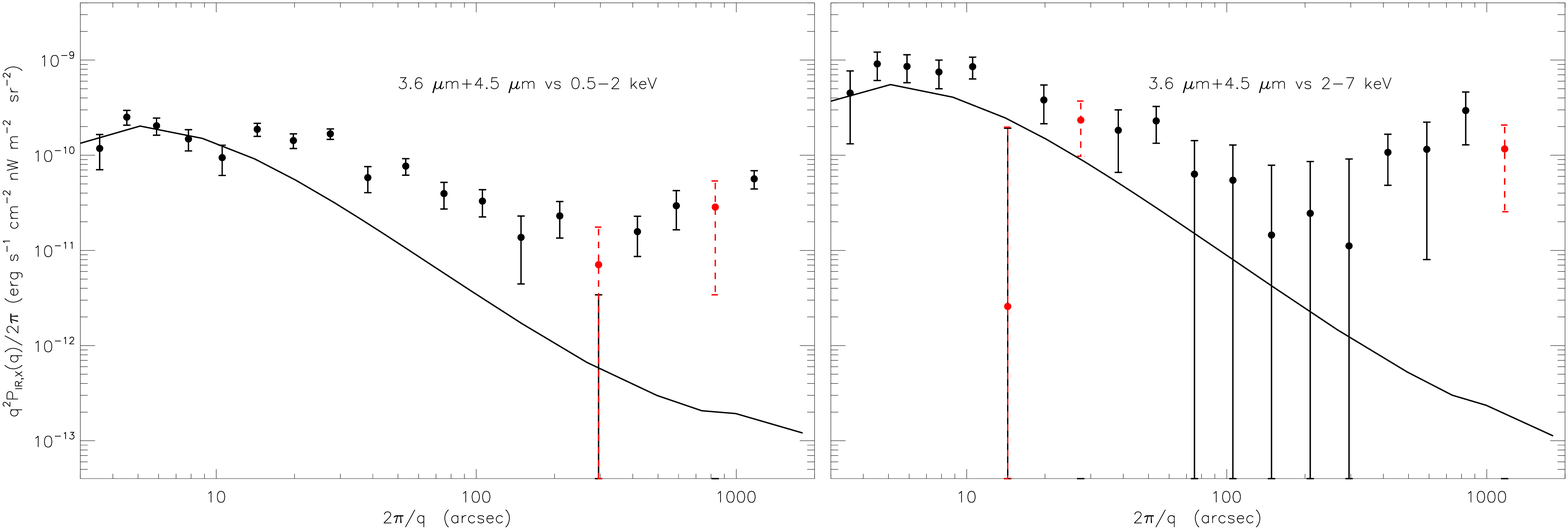}

\caption{\label{fig:cps} Combined CIB-CXB cross-power spectra ($black~filled~ circles$). 
$Top$, $center$ $panels:$
3.6~$\mu$m vs. 0.5-2~keV, 4.5~$\mu$m vs .0.5-2~keV, 3.6~$\mu$m vs. 2-7~keV and 4.5~$\mu$m vs. 2-7~keV.
We over-plot our reconstruction of known $z<6$ populations: $dashed~line$ for AGN, dot-dashed line for star-forming galaxies, $dotted~lines$ for hot gas in clusters and solid lines for the sum.
$Bottom~panels:$ Cross-power of combined CIB (3.6+4.5~$\mu$m) vs. $(left)$ soft X-ray (0.5-2~keV) and $(right)$ hard X-ray (2-7~keV) bands. }
\end{figure*}

After masking detected-source pixels, we study the surface brightness fluctuation field, $\delta$F and compute its Fourier Transform $\Delta(q)$ using FFT with $q$ as the angular frequency. The power spectrum in a single band $n$ is $P_n(q)=\langle|\Delta(q)|^2\rangle$, where the average is taken over the interval [$q$, $q + dq$] and its error is $\sigma_{P_{n}}(q)=P_{n}(q)/\sqrt{N_q}$, where $N_q$ is the number of independent 
Fourier elements adopted in the analysis, and $\sqrt{q^2 P_n(q)/2\pi}$ is the typical rms fluctuation in the flux on a scale with wavelength 2$\pi$/$q$. In this study, the masked pixels 
occupy 30-40\% of the maps (see Table \ref{tbl-1}), which allows for a robust FFT analysis \citep{k12}. In order to determine the intensity and structure of the joint fluctuations for every pair of independent photometric bands, we estimated the cross-power spectrum using:
$P_{m,n}(q) = \Delta_m(q)\Delta_{n}(q)^*= Re_m(q)Re_n(q) + Im_m(q)Im_n(q)$, where $Re$ and $Im$ refer to the real and imaginary parts of the Fourier transform. 
The errors are:
\begin{equation}
\sigma_{P_{m,n}}(q)=\sqrt{P_m(q)P_n(q)/N_q}.
\label{eq:eq1}
\end{equation}
All the IR maps have been clipped 
at the same shot noise level to combine the 4 fields to reduce cosmic and sample variance. We performed Fourier analysis in the four fields 
listed in Table \ref{tbl-1} and averaged the cross-power spectra by weighting with their errors. 
In order to combine signals, all the fields of different geometries were binned in Fourier space to give power at identical $q$.
The stacked cross-power is computed by averaging $\langle\Delta_m\Delta_n^*\rangle$
and its variance: $\sigma^2_{P_{(n,m)}}(q)=\frac{1}{\sum_{i=1}^4 (\sigma^i_{P_{m,n}}(q))^{-2}}$.

\subsection{Systematic effects}

One concern is the possibility of either random or 
spurious cross-correlation in the data. 
In order to evaluate this, we cross-correlated the IR fluctuations with the X-ray noise maps obtained by
subtracting maps of even events (A) from maps of odd events (B). These A-B maps 
contain only random noise/artifacts \citep[see][]{cap2013}. 
Figure \ref{fig1} shows that this 
cross-power is 
very low at scales larger than $\sim20$~arcsec,
for every individual field and the combined fields; 
the larger deviations below 20~arcsec likely occur because the masked regions have these dimensions. 
The noise floor is about one dex higher 
for correlations with
the hard X-ray band compared to the soft X-ray band, 
due to worse statistics.
We conclude that in our maps noise and instrumental effects are uncorrelated on scales above an arcminute. 


\section{Results}\label{sect3}

For each survey field we evaluated the cross-power spectrum in four possible IR and X-ray band pairs, as shown in
the lower panels of Figure \ref{fig1}. 
While  in the individual fields there appears, at scales  $>$20\arcsec, a significantly positive signal (Table \ref{tbl-2}) when cross-correlating CIB with [0.5-2] keV, its significance improves dramatically when we combine all four fields, as shown in Figure~\ref{fig:cps}. 

On the angular scales sampled here, the correlations of
3.6~$\mu$m vs. 0.5-2~keV, 4.5~$\mu$m vs. 0.5-2~keV and 4.5~$\mu$m vs. 2-7~keV show a positive cross-power at the 6$\sigma$, 7.8$\sigma$ and 2.1$\sigma$ confidence level, respectively; 
for 3.6$\mu$m vs. [2-7]~keV the signal is positive but consistent with zero at 1$\sigma$ (cf. \citealt{mw16}).
The novelty of our analysis lies in the combination of deeper fields, at both IR and X-ray wavelengths, over a much larger area.  
The cross-power amplitude for the four combined fields on large scales (20\arcsec-1500\arcsec), reported in Table~\ref{tbl-2}, is significant at  5$\sigma$ and 6.3$\sigma$ for 3.6~$\mu$m vs. 0.5-2~keV and 4.5~$\mu$m vs. 0.5-2~keV, respectively; there is no significant cross-power between either IR channel and the hard band X-rays. 
In the bottom panel of Fig. \ref{fig:cps} the broad 3.6+4.5 $\mu$m vs 0.5-2~keV and 2-7~keV cross-powers were obtained by averaging the measurements of P(q) in the sub-bands. 
At 5\arcsec-1500\arcsec, the CIB correlates with the soft band at $>$10$\sigma$ and with the hard band at $\sim$3.5$\sigma$. 
 The correlation above 20\arcsec is significant at the $>$10$\sigma$ and ~2.5$\sigma$ level, for soft and hard X-ray bands, respectively. 
 
As a reference, on order to evaluate the level of correlation between the two band pairs we evaluated the level of coherence of the fluctuations $\mathcal{C}\sim$0.14-0.20 for either IR channel versus the soft band.

\section{Discussion}\label{sect4}

The significant cross-correlation signal arises from a population of sources that emit both in IR and X-rays or share the same environment. Known sources of extragalactic X-rays include i) normal galaxies, ii) AGN and iii) hot gas in clusters and groups. In the following, we use the cross-power reconstruction of Helgason et al. (2014) with some improvements.


Galaxies contain high- and low-mass X-ray binaries whose X-ray luminosities scale with star formation rate and stellar mass respectively \citep[e.g.][]{basu}:
\begin{equation} \label{relation}
  L_{\rm X} = \alpha {\rm SFR} (1+z)^\gamma + \beta M_\star (1+z)^\delta ~, 
\end{equation}
where $\alpha$, $\beta$, $\gamma$ and $\delta$ are parameters for which we adopt the values measured by \citet{leh16} in both the soft and hard band, and include the intrinsic scatter in the relation.
For the underlying galaxy population we use a semi-analytic galaxy formation model based on the Millennium simulation \citep{Henriques15}, which is in good agreement with the observed star formation history and stellar mass function as a function of redshift.


We use the IR brightness and a projected position given by the model light cones to create a model image. The brightness distribution is also in a good agreement with observed galaxy counts and luminosity functions. We assign an X-ray brightness according to Equation (2) to the same image position based on the physical properties and the luminosity distance of the galaxy. To mimic the source masking, we eliminate all sources with IR magnitude brighter than m$_{AB}$=24.8 and calculate the angular power spectrum of the remaining sources in the same way as described in Section 3. The magnitude limit is tuned to match the shot noise level in the IR auto power spectrum, which is known to be galaxy-dominated. On large scales however, the IR auto power spectrum is lower than measurements and is in agreement with Helgason et al. (2012).

\begin{figure*}
\includegraphics[width=\textwidth]{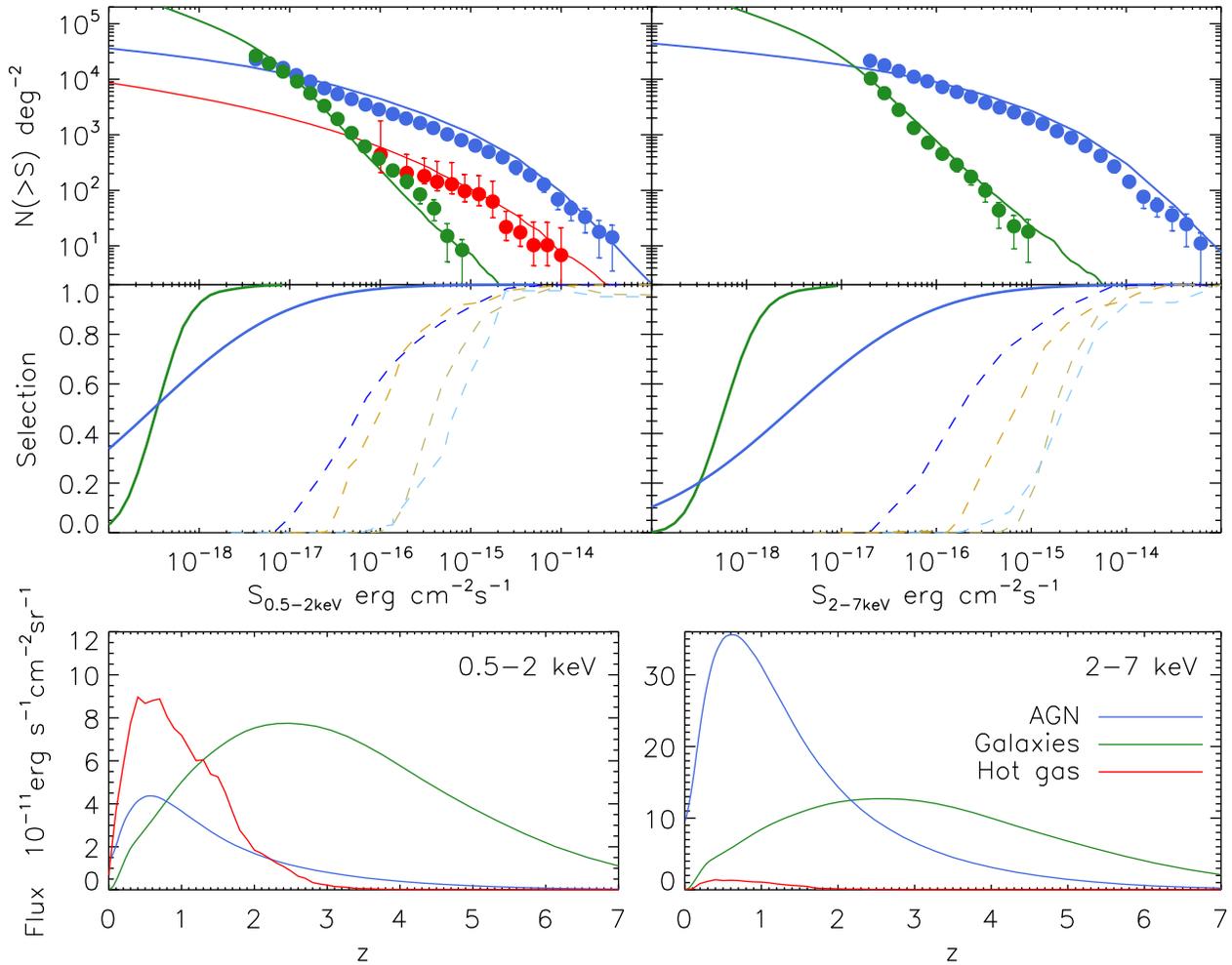}
\caption{\label{fig3}{\it Top:} X-ray source counts for AGN, galaxies (Luo et al. 2017) and clusters/groups (Finoguenov et al. 2015) shown in blue, green, red respectively in soft X-ray (0.5--2~keV, {\it left}) and hard X-ray (2--7~keV, {\it right}). Our models are shown as solid lines for comparison. {\it Middle:} The fraction of X-ray detected and masked sources as a function of flux in our four fields are shown as dashed lines in the same color scheme as Figure \ref{fig1}. Solid lines show the modeled AGN (blue) and galaxies (green) remaining after applying the joint X-ray/IR mask. {\it Bottom:} Remaining unresolved CXB emission from cosmic populations contributing to the cross-power (same color scheme as above).}
\end{figure*}

For the AGN contribution, we adopt the population model of \citet{gilli07} in X-rays and Helgason et al. (2014) in IR. The extent to which AGN are removed by the joint IR/X-ray mask is estimated using empirical X-ray-to-optical relations (\citealt[; for details see]{civ12} \citealp{h14}). The fraction of removed sources as a function of brightness, referred to as the selection function, is shown in Fig. \ref{fig3}. The extended tail of the AGN selection is due to the large intrinsic scatter in the X-ray to IR relation for AGN and is the most uncertain factor in our calculation. The shot noise however gives us a constraint on how large this scatter can actually be. Interestingly, in order to simultaneously match the amplitude of the small scale cross-power the 0.5-2 keV and  2-7 keV band we need to assume an extremely hard spectral slope ($\Gamma=0.5$), possibly implicating heavily obscured AGN responsible for the power on small scales.

Hot X-ray emitting gas in groups and clusters of galaxies spatially correlates with IR emitting sources sharing the same environments. We adopt the hot gas modeling of \citet{h14} (Sec. 5.1.3) which uses the mass and extent of hot gas from the same semi-analytic model used for galaxies above \citep{guo11, Henriques15}. We assume a beta-model density profile of hot gas in halos emitting with a simple Brehmmstrahlung spectrum determined by the gas temperature. Finally, we tune the average gas mass in halos (by a factor of 0.35) to match observed X-ray group/cluster counts (see Fig. \ref{fig3}). To mimic the masking of groups in our several fields we adopt the 50\% group detection completeness level of the ECDF-S (Finoguenov et al. 2015).
 
Figure~\ref{fig:cps} shows the contribution from galaxies, AGN and clusters to the cross-power of the unresolved IR and X-ray sources. All modeled cross-power spectra are multiplied by the Chandra beam for which we use the analytic profile given in \citep{kolo}. The evolution of the CXB production rate of the reconstructed populations is shown in Fig. \ref{fig3}.
The soft band X-ray flux expected from summing known but unresolved populations (Fig.~\ref{fig3}) is $\sim$2.6$\times$10$^{-13}$~erg~cm$^{-2}$~s$^{-1}$~deg$^{-2}$, which is $\sim$30\% of the still unresolved CXB flux or 2.5\% of the total CXB flux \citep{cap2017}, so unknown population(s) contribute $\lesssim$7$\times$10$^{-13}$ erg cm$^{-2}$ s$^{-1}$ deg$^{-2}$.  \\

We detect a cross-power signal that is well explained on small scales ($<20$~arcsec) by unresolved, known sources (galaxies, AGN, clusters) but on larger scales in excess at $\sim$5$\sigma$ of those populations (Fig.~\ref{fig:cps}).
Scaling up the contribution from known sources to match the large scale cross-power  from clustering would strongly over-predict the signal on small scales from shot-noise. 

We considered possible explanations for the observed excess cross-power on large scales.
\citet{yue13} proposed that the observed CIB-CXB coherence could be explained by a population of Compton-thick DCBHs at z$>$12. 
However, by $z\sim$10 their model would already produce an accreted mass-density of BHs greater than the observed value locally \citep{h16}. 
Perhaps by tuning the parameters, their model could satisfy this integral constraint while still allowing massive rapidly growing DCBHs to account for much of the observed excess.

\citet{k16} recently proposed that primordial BHs, if they exist in sufficient numbers to account for the entire dark matter content of the universe, would produce the extra small scale power in matter fluctuations to explain the measured Spitzer-based CIB fluctuations.
 
In that case, accreting BHs like those observed with LIGO \citep{ligo}, with masses
$\sim20-60 M_{\odot}$, could naturally produce part or all the observed excess.

\citet{c12} and \citet{zemcov} suggested that "orphan" stars at $z\sim1-5$  in a diffuse 
 intra-halo light  could fully explain the detected excess CIB fluctuation.
 Our measurement of the CIB vs CXB coherence means that 
 intra-halo light could produce most of the CIB excess only if
 a substantial fraction  (larger than that observed in galaxies) of the orphan stars are X-ray binaries or pulsars or share the same environment with hot X-ray emitting gas.
 
 Finally, some fraction of the CIB excess may arise from Galactic light scattered by interstellar dust. This Diffuse Galactic Light is very faint at
3.6 and 4.5~$\mu$m, and is generally estimated through extrapolation from, or cross correlation with, much brighter interstellar emission at other wavelengths \citep[e.g.,][]{k2005, a10, zemcov, m11, seo}. Galactic X-rays also scatter in the diffuse ISM \citep{molaro} and thus might correlate with the IR.
However, X-ray scattering is predominantly a small angle phenomenon, dropping sharply with increasing angular scale \citep{sm, vs}, 
so the X-ray sources would have to be within $\sim$1000$\arcsec$ of the survey fields. At the high latitudes of the deep surveys, there are very few Galactic sources \citep{lehmer} and we estimate the flux of such a component to $\sim$10 below our fluctuations.. 

Forthcoming missions like $Euclid$, $WFIRST$, $JWST$, $eROSITA$ and $Athena$ offer powerful new ways to address the true nature of the cross CIB-CXB fluctuations

\acknowledgments
NC thanks Yale University for the YCAA Prize Postdoctoral Fellowship. 
We  acknowledge NASA ADAP grant NNX16AF29G,  Chandra SAO grant AR6-17017B,
 Chandra SAO grant GO5-16150A and NASA/12-EUCLID11-0003. We thank the anonymous referee
 for comments/suggestions.


{\it Facilities:} \facility{CXO (ACIS) \facility{Spitzer (IRAC)}}.

\clearpage

\end{document}